\documentclass{tlp}

\usepackage{learning-macros}
\usepackage{logic}
\usepackage{listings}
\usepackage{subfigure}
\usepackage{paralist}
\usepackage{url}
\usepackage{fancyvrb}
\usepackage{multicol}
\usepackage{tikz}

\makeatletter
\newif\if@restonecol
\makeatother

\usepackage{algorithm2e}        
\usetikzlibrary{positioning}
\usetikzlibrary{shapes}

\newtheorem{definition}{Definition} 
\newtheorem{theorem}{Theorem}

\begin{document}
\bibliographystyle{acmtrans}

\long\def\comment#1{}

\title{Normative Design using Inductive Learning}

\author[D. Corapi, M. De Vos, J. Padget, A. Russo and K. Satoh]
{DOMENICO CORAPI, ALESSANDRA RUSSO\\
Department of Computing\\
Imperial College London\\
180 Queen's Gate, SW7 2AZ\\
London, UK   \\
E-mail: \{d.corapi,a.russo\}@ic.ac.uk
\and
MARINA DE VOS, JULIAN PADGET \\
Department of Computing\\
University of Bath, BA2 7AY \\
Bath, UK   \\
E-mail: \{mdv,jap\}@cs.bath.ac.uk
\and
KEN SATOH\\
Principles of Informatics Research Division\\
National Institute of Informatics\\
Chiyoda-ku, 2-1-2, Hitotsubashi\\ 
Tokyo 101-8430, Japan\\
E-mail: ksatoh@nii.ac.jp
}

\pagerange{\pageref{firstpage}--\pageref{lastpage}}
\volume{\textbf{10} (3):}
\jdate{January 2011}
\setcounter{page}{1}
\pubyear{2011}

\maketitle
\label{firstpage}

\begin{abstract}

In this paper we propose a use-case driven iterative design methodology for normative frameworks, 
also called virtual institutions, which are used to govern open systems.
Our computational model 
represents the normative framework as a logic program under answer set semantics (ASP). 
By means of an inductive logic programming (ILP) approach, implemented using ASP, it is possible to synthesise 
new rules and revise existing ones. The learning mechanism is guided by the designer who describes the desired properties of the 
framework through use cases, comprising
  \begin{inparaenum}[(i)]
    \item event traces that capture possible scenarios, and
    \item a state that describes the desired outcome.
  \end{inparaenum}
  The learning process then proposes additional rules, or changes to
  current rules, to satisfy the constraints expressed in the use
  cases.  Thus, the contribution of this paper is a process for the
  elaboration and revision of a normative framework by means of a
  semi-automatic and iterative process driven from specifications of
  (un)desirable behaviour. The process integrates a novel and general
  methodology for theory revision based on ASP.

  \end{abstract}

\section{Introduction}

Norms and regulations play an important role in the governance of
human society. Social rules such as laws, conventions and contracts
prescribe and regulate our behaviour. 
By providing the means to describe and reason about norms in a
computational context, normative frameworks (also called institutions
or virtual organisations) may be applied to software systems. Normative frameworks allow
for automated reasoning about the consequences of socially acceptable
and unacceptable behaviour by monitoring the permissions, empowerment
and obligations of the participants and generating violations when
norms are not followed.

Just as legislators, and societies, find inconsistencies in their rules (or
conventions), so too may designers of normative frameworks. 
The details of the specification
makes it relatively easy to miss crucial operations needed to help or
inhibit intended behaviour.  To make an analogy with software
engineering, this characterises the gap between requirements and
implementation and what we describe here can be seen as an
automated mechanism to support the validation of normative frameworks, 
coupled with regression testing.

The contribution of the work is twofold. Firstly, we show how inductive logic
programming (ILP) can be used to fill gaps in the rules of an existing normative framework. 
The designer normally develops a system with a certain behaviour in mind. This intended behaviour
can be captured in {\em use cases} which comprise two components: a description of a scenario 
and the expected outcome when executing the scenario. 
Use cases are added to the program to validate
the existence of an answer set. Failure to solve the program indicates that the specification does not yield
the intended behaviour. In this case, the program and the failing use case(s) are given to an inductive learning
tool, which will then return suggestions for improving the normative specification such that the use cases are satisfied.
Secondly, we present a novel integrated methodology for theory revision that can be used 
to revise a logic program under the answer set semantics (ASP) and supports
the development process by associating answer sets (that can be used
for debugging purposes) to proposed revisions.  Due to the non-monotonic nature of ASP, the designer can provide the essential parts of the use case creating a template rather that a fully specified description. 
The revision mechanism is general and can be applied to other domains.
We demonstrate the methodology through a case study showing the iterative revision process.

The paper is organised as follows. Section~\ref{sec:normative}
presents some background material on the normative framework, while
Section~\ref{sec:learning} introduces the ILP setting
used in our proposed approach. 
Section~\ref{sec:learning normative rules} illustrates the methodology and how
the revision task can be formulated into an ILP problem.
We illustrate the flexibility and
expressiveness of our approach through 
specifications of a reciprocal file sharing normative system.
Section~\ref{sec:details} discusses the details of the revision mechanism
and the learning system.
Section~\ref{sec:related} relates our approach to existing work. We conclude with a summary and remarks on future work.







\section{Normative Frameworks}\label{sec:normative}


The essential idea of normative frameworks is a (consistent)
collection of rules whose purpose is to describe {\em a principle of
right action binding upon the members of a group and serving to
guide, control, or regulate proper and acceptable
behaviour [Merriam-Webster dictionary]}.  These rules may be stated in
terms of events, specifically the events that matter for the
functioning of the normative framework.  

\begin{figure}[!t]
\subfigure[]{\label{first}    \label{formal model} 
\begin{minipage}[b]{0.35\textwidth}%
\footnotesize\raggedright
\renewcommand{\union}{\cup}
\renewcommand{\intersection}{\cap}
{$\mathcal{N} = \basicinsttuple$}, where
\begin{enumerate}
\item\label{fm facts} $\factprops = \powers \union \perms \union \obls
  \union \domains$ 
\item\label{fm gen} {$\evgens: \expressions \times \events \rightarrow
    \powerset{\instevents}$}
\item\label{fm con} {$\consequences : \expressions \times \events \to
    \powerset{\factprops} \times \powerset{\factprops}$} where\\ $C(X,e)
  = (\conseqinit{\anexpression}{e},\conseqterm{\anexpression}{e})$ where
\begin{enumerate}[(i)]
\item {$\conseqinit{\anexpression}{e}$} initiates a fluent
\item {$\conseqterm{\anexpression}{e}$} terminates a fluent
\end{enumerate}
\item\label{fm events} {$\events = \obsevents \cup \instevents$ \\with
    $\instevents = \instactions \cup \violevents$}
\item\label{fm init} {$\initials$}
\item\label{fm sf} State Formula: {$\expressions =
    \powerset{\factprops \union \lnot \factprops}$}
\end{enumerate}
\end{minipage}
}%
\subfigure[]{\label{second}     \label{translator}  \label{fig2}
\begin{minipage}[b]{0.6\textwidth}%
\footnotesize
\begin{align}
p \in \factprops  \Leftrightarrow & \asp{ifluent(p)}.  \label{ifluent}\\[-3pt]
e \in \events  \Leftrightarrow & \asp{event(e)}.\label{event}\\[-3pt]
e \in \obsevents \Leftrightarrow & \asp{evtype(e,obs)}. \\[-3pt]
e \in \instactions \Leftrightarrow & \asp{evtype(e,act)}.\\[-3pt]
e \in \violevents \Leftrightarrow & \asp{evtype(e,viol)}. \\[-3pt]
\conseqinit{\anexpression}{e} = P \Leftrightarrow & \Forall{p \in P}{\asp{initiated(p,T)} \notag\\[-3pt] & \asp{\leftarrow} \ occurred(e,I),\trans{\anexpression}{T}}.\label{initiated}\\[-6pt]
\conseqterm{\anexpression}{e} = P \Leftrightarrow & \Forall{p \in P}{\asp{terminated(p,T)} \notag\\[-3pt] & \asp{\leftarrow} \ occurred(e,I),\trans{\anexpression}{T}}.\label{terminated}\\[-3pt]
\evgens(\anexpression,e) = E \Leftrightarrow & g \in E, \notag\\[-3pt]
&\asp{occurred(g,T)\asp{\leftarrow} occurred(e,T),}\notag\\[-3pt]
  &\asp{holdsat(pow(e),I),}\trans{\anexpression}{T}.\label{occurred}\\[-3pt]
p \in \initials  \Leftrightarrow & \asp{holdsat(p,i00)}.
\end{align}
\end{minipage}
}
\caption{\subref{first} Formal specification of the normative framework and \subref{second} translation of normative framework specific rules into \AnsProlog}
\end{figure}

\subsection{Formal Model}

The formalization of the above may be defined as
conditional operations on a set of terms that represent the normative
state.  
To provide the context for this paper, we give an outline of a formal
event-based model for the specification of normative frameworks that
captures all the essential properties, namely {\em empowerment}, {\em permission},
{\em obligation} and {\em violation}.  We adopt the formalisation from 
\cite{cdvp06a}, summarized in Figure~\ref{formal model}, because of its straightforward mapping to answer set
programming. 

The essential elements of the normative framework are events ($\events$), which bring about changes in state, and
fluents ($\fluents$), which characterise the state at a given instant.
The function of the framework is to define the interplay between these
concepts over time, in order to capture the evolution of a particular
institution through the interaction of its participants.  We
distinguish two kinds of events: normative events ($\instevents$),
that are the events defined by the framework, and exogenous events
($\exevents$), some of whose occurrence
may trigger normative events in a direct reflection of ``counts-as''
\cite{jones+sergot:1996}, and others that are of no relevance to this particular framework. 
Normative events are further partitioned into
normative actions ($\instactions$) that denote changes in normative
state and violation events ($\violevents$), that signal the occurrence
of violations.  Violations may arise either from explicit generation,
(i.e. from the occurrence of a non-permitted event), or from the non-fulfilment of an obligation.  We also distinguish two kinds of fluents: {\em
  normative fluents} that denote normative properties of the state
such as {\em permissions} ($\perms$), {\em powers} ($\powers$) and obligations
($\obls$), and {\em domain fluents\/} ($\domains$) that correspond to
properties specific to a particular normative framework. 
A normative state is represented
by the fluents that hold true in this state.  Fluents that are not
present are considered to be false. 
Conditions on a state ($\expressions$) are expressed by a set of fluents that should be true or false. 
When the creation event occurs, the normative state is initialised with
the fluents specified in $\initials$.

Changes in a normative state are achieved through the definition of two
relations:
\begin{inparaenum}[(i)]
\item the generation relation ($\generates$), which implements counts-as by
  specifying how the occurrence of one (exogenous or normative) event
  generates another (normative) event, subject to the empowerment of
  the actor and the conditions on the state, and
\item the consequence relation ($\consequences$), which specifies the initiation
  and termination of fluents, subject to the performance of some action
  in a state matching some condition. \end{inparaenum}

The semantics of a normative framework is defined over a sequence,
called a {\em trace}, of exogenous events.  Starting from the initial
state, each exogenous event is responsible for a state change, through
initiation and termination of fluents. This is achieved by a
three-step process:
\begin{inparaenum}[(i)]
\item the transitive closure of $\generates$ with respect to a given
  exogenous event determines all the generated (normative) events,
\item to this all violations of non-permitted events and non-fulfilled 
   obligations are added, giving the set of all events whose
  consequences determine the new state,
\item the application of $\consequences$ to this set of events
  identifies all fluents that are initiated and terminated with
  respect to the current state, so determining the next state.
\end{inparaenum}
For each trace, we can therefore compute a sequence of states that
constitutes the model of the normative framework for that trace.  This
process is realised as a computational model through answer set
programming (see Section~\ref{sec:modelling normative frameworks}) and it is
this representation that is used in the learning process
described in Section~\ref{sec:learning normative rules}. A detailed example
of the formal model of an institution can be found in \cite{cdvp06a}.

\subsection{Computational Model}\label{sec:modelling normative frameworks}
The formal model described above can be translated
into an equivalent computational model using answer set programming (ASP) \cite{gellif91} with \AnsProlog as the
implementation language.
\AnsProlog is a knowledge representation language that allows the programmer to describe 
a problem and the requirements on the solutions
in an intuitive way, rather than the algorithm to find the solutions to
the problem. For our mapping, we followed the naming convention used in
the event calculus \cite{KS86} and action languages \cite{gellif_act98}.

 \begin{sloppypar}
The basic components of the language are atoms, elements that can be
assigned a truth value. An atom can be negated using {\em negation as failure}. {\em Literals} are atoms $a$ or negated atoms $\notf{a}$. We say that $\notf{a}$ is true if
we cannot find evidence supporting the truth of $a$.
Atoms and literals are used to create rules of the general form:
\ $a \leftarrow  b_{1}, ..., b_{m}, \notf{c_{1}}, ..., \notf{c_{n}}$, where $a$, $b_{i}$ and $c_{j}$ are atoms. Intuitively, this means
{\it if all atoms $b_{i}$ are known/true and no atom $c_{j}$ is known/true, then $a$ must be known/true}. We refer to $a$ as the head and
$b_{1}, ..., b_{m}, \notf{c_{1}}, ..., \notf{c_{n}}$ as the body of the rule. Rules with empty body are  called {\em facts}. Rules with empty head are referred to 
as {\em constraints}, indicating that no solution should be able to satisfy the body.
A {\em (normal) program (or theory)} is a conjunction of rules and is also denoted by a set of rules.
The semantics of \AnsProlog is defined in terms of {\em answer sets}, 
i.e. assignments of true and false to all atoms in the program 
that satisfy the rules in a
minimal and consistent fashion.  A program may have zero or more answer sets,
each corresponding to a solution.
 \end{sloppypar}


The mapping of a normative framework consists of three parts: a {\em base
component} which is independent of the framework being modelled, the {\em
time component} and the {\em framework specific component}. The independent
component deals with inertia of the fluents, the generation of
violation events of non-permitted actions and of unfulfilled
obligations. The time component defines the predicates for time and is
responsible for generating a single observed event at every time
instance.  The mapping uses the following atoms: \asp{ifluent(p)} to
identify fluents, \asp{evtype(e,t)} to describe the type of an event,
\asp{event(e)} to denote the events, \asp{instant(i)} for time
instances, \asp{final(i)} for the last time instance,
\asp{next(i1,i2)} to establish time ordering, \asp{occurred(e,i)} to
indicate that the (normative) event happened at time $i$,
\asp{observed(e,i)} that the (exogenous) event was observed at 
time $i$, \asp{holdsat(p,i)} to state that the normative fluent $p$ holds
at $i$, and finally \asp{initiated(p,i)} and \asp{terminated(p,i)} for
fluents that are initiated and terminated at $i$.  
Note that exogenous
events are always empowered, so that observed events are always
occurred events, but that normative events are not, so their
occurrence is conditional on their empowerment.
Figure~\ref{translator} provides the framework specific translation rules,
including the
definition of all the fluents and events as facts.  We translate
expressions into \AnsProlog rule bodies as conjunctions of literals
using negation as failure for negated expressions.

The translation of the formal model is augmented with a trace program,
specifying the length of traces that the designer is interested in and
rules to ensure that, all but the final time instance, is associated
with exactly one exogenous event. Specific occurrences of events can be specified
as facts (e.g. \asp{observed(event,instance)}). We refer to a complete trace when all
exogenous events for a giving time interval are specified. If a trace is incomplete
when the model needs to determine the missing exogenous events. While not discussed in this paper, both the normative
framework and the learning tool can deal with both types of traces. 
When the model is supplemented with the \AnsProlog specification
of a complete trace, 
we obtain a single answer set corresponding to the model matching
the trace\footnote{The structure of the program (the stratified base part and 
observed events as facts), guarantees that the program has exactly one answer set. See \cite{cliffe:2007} for further details
and proofs.}.
In this case the complexity of computing the answer set is linear
with respect to the number of time instance being modelled. This result
can easily be derived from the structure of the program.
Of course, in the absence of a complete trace, the complexity is 
NP-complete as the traces composed of all possible combinations 
of missing exogenous events are computed. See \cite{cliffe:2007} for further details
and proofs.

%

\section{Learning}
\label{sec:learning}
 \begin{sloppypar}
Inductive Logic Programming (ILP) \cite{Muggleton95}  is a machine learning technique concerned with the induction of logic theories that generalise (positive and negative) examples with respect to a prior background knowledge. 
For example, from the observations (properties in this paper) $P_{fly} = \{fly(a), fly(b), \notf fly(c)\}$ and a background knowledge containing the two facts $bird(a)$ and  $bird(b)$, we can generalise the concept $fly(X) \ifl bird(X)$. In non-trivial problems it is crucial to define the space of possible solutions accurately. Target theories are within a space defined by a {\em language bias}, that can be expressed using the notion of mode declaration \cite{Muggleton95}. 
 \end{sloppypar}

\begin{definition}
\label{mode declaration}
A {\em mode declaration} is either a {\em head declaration}, written  $modeh(s)$, or a {\em body declaration}, written $modeb(s)$, where $s$ is a  {\em schema}. A schema is a ground literal containing special terms called {\em placemarkers}. A {\em placemarker} is either `$+type$', `$-type$' or `$\#type$' where {\em type} denotes the type of the placemarker and the three symbols `$+$', `$-$' and `$\#$' indicate that the placemarker is an input, an output and a constant respectively.  
\end{definition}

\begin{sloppypar}
In the previous example a possible language bias would be expressed by three mode declarations in $M_{fly}$: $modeh(fly(+animal))$, $modeb(bird(+animal))$ and $modeb(penguin(+animal))$.
\end{sloppypar}

A rule $h \ifl b_{1}, ..., b_{n}$ is {\em compatible} with a set $M$ of mode declarations iff  (a) $h$ is the schema of a head declaration in $M$ and $b_{i}$ are the schemas of body declarations in $M$ where every input and output placemarkers are replaced by variables, and constant placemarkers are replaced by constants; (b) every input variable in any atom $b_{i}$ is either an input variable in $h$ or an output variable in some $b_{j}, j < i$; and (c) all variables and constants are of the corresponding type (enforced by implicit conditions in the body of the rules). From a user perspective, mode declarations establish how rules in the final hypotheses are structured, defining literals that can be used in the head and in the body of a well-formed hypothesis. $s(M)$ is the set of all the rules compatible with $M$.

\begin{definition}
\label{ilp}
An {\em ILP task} is a tuple $\langle P, B, M\rangle$  where $P$ is a set of conjunctions of literals,  called {\em properties},  $B$ is a normal program, called {\em background theory}, and $M$ is a set of mode declarations. A theory $H$, called {\em hypothesis}, is an inductive solution for the task $\langle P, B, M\rangle$, if  
(i)~$H \subseteq s(M)$,  and (ii)~$P$ is true in all the answer sets of $B \cup H$.
\end{definition}


Our approach for incremental development of a normative system supports the synthesis of new rules and revision of existing one from given use-cases. We are therefore interested in the task of Theory Revision (TR). As discussed in \cite{Corapi2009}, non-monotonic inductive logic programming can be used to revise an existing theory.
The key notion is that of {\em minimal revision}.  In general, a TR system is biased towards  the computation of theories that are similar to a given revisable theory.  
Our revision algorithm uses a measure of minimality similar to that proposed in  \cite{audrey2}, and defined in terms of {\em number of revision operations} required to transform one theory into another.

\begin{definition}
\label{tr1}
Let $T'$ and $T$ be normal logic programs. A revision transformation $r$ is such that $r(T)=T '$, and $T'$ is obtained from $T$ by deleting a rule, 
adding a fact, adding a condition to a rule in $T$ or deleting a condition from a rule in $T$. $T'$ is a revision of $T$  with distance $c(T, T') = n$ iff $T' = r^{n}(T)$ and there is no $m < n$ such that $T' = r^{m}(T)$.
%
\end{definition}

For example, given the theory $T_{fly} = \{fly(X) \ifl bird(X)\}$, $T_{fly}' = \{fly(X) \ifl bird(X), not\ penguin(X)\}$ is a revision of $T$ with distance $1$.
Note that, although we refer to Definition~\ref{tr1}, it is also possible to weight revisions differently or introduce different transformations. 

\begin{definition}
\label{tr}
A {\em TR} task is a tuple $\langle P, B, T, M\rangle$  where $P$ is a set of conjunctions of literals, called {\em properties},  $B$ is a normal program, called {\em background theory}, $T \subseteq s(M)$ is a normal program, called {\em revisable theory}, and $M$ is a set of mode declarations. 
The theory $T'$, called {\em revised theory}, is a {\em TR solution} for the task $\langle P, B, T, M\rangle$ with distance $c(T, T')$, iff  
(i) $T' \subseteq s(M)$, (ii)
$P$ is true in all the answer sets of $B \cup T'$, (iii) if a theory $S$ exists that satisfies conditions (i) and (ii) then $c(T, S) \geq c(T, T')$, (i.e. minimal revision). 
\end{definition}

For example, let $B_{fly} = \{animal(X). \hspace{2mm}bird(X). \hspace{2mm}penguin(c).\}$,  $T_{fly}$, $P_{fly}$ and $M_{fly}$ as in the previous examples. $T_{fly}'$ is a TR solution for the task $\langle P_{fly}, B_{fly}, T_{fly}, M_{fly}\rangle$ with distance $1$.
The main difference with the ILP task given in Definition~\ref{ilp} is the availability of an initial revisable theory and the consequent bias, as discussed in more detail in the following sections. 



\section{Revising Normative Rules}\label{sec:learning normative rules}

\subsection{Methodology}

Use cases represent instances of executions that are known to the
designer and that drive the elaboration of a normative system. If
the current formalisation of a normative system does not match the intended
behaviour in the use cases then the formalisation is not complete
or is incorrect, and an extension or revision is required.


Each use case $u \in U$ is a tuple $\langle T, 
O\rangle$ where $T$, a {\em trace\/}, specifies a set of
exogenous events (\asp{observed(e,t)}), 
and $O$ is a set of \asp{holdsat} and \asp{occurred} literals that represent the {\em expected output} of the use case.
Given a set $U$ of use cases, $T_{U}$ and $O_{U}$ denote, respectively, the set of all the traces and expected outputs in all the use cases in $U\/$. The time points of the different use cases relate to different instances of executions of the normative system to avoid the effect of events in one use case affecting the fluents of another use case. 
The use cases can, but do {\em not} have to, be complete traces (i.e. an event for each time instance) and expected output can contain positive as well as negative literals. 

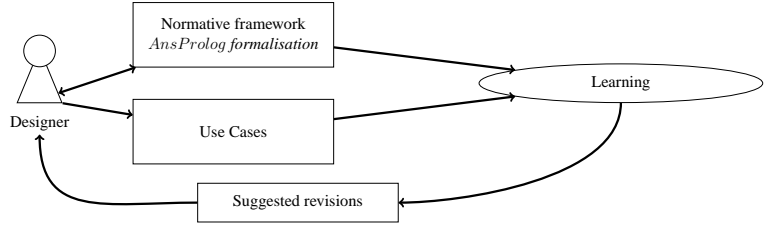
\begin{figure}[!t]
\begin{center}
\resizebox{0.8\columnwidth}{!}{
\begin{tikzpicture}
\node[rotate=90,isosceles triangle,draw,thick](body){\rule{0cm}{1cm}};
\draw(body)+(0,1.5)node[circle,thick,draw,fill=white](head){\rule{0cm}{0.75cm}};
\draw(body)+(0,-0.75)node(e){\Large Designer};
\draw(body)+(6,2)node[rectangle,draw,minimum width=1cm,minimum height=2cm](a){\begin{minipage}{6cm}\Large\centering
Normative framework\\
\AnsProlog {\em formalisation}
\end{minipage}};

\draw(body)+(6,-1.0)node[rectangle,draw,minimum width=3cm,minimum height=2cm](b){\begin{minipage}{6cm}\Large\centering
Use Cases
\end{minipage}};

\draw(body)+(18,0.5)node[ellipse,draw,minimum width=2cm,minimum height=1.2cm](c){\begin{minipage}{6cm}\Large\centering
Learning
\end{minipage}};

\draw(body)+(8,-3.2)node[rectangle,draw,minimum width=3cm,minimum height=1.2cm](d){\begin{minipage}{6cm}\Large\centering
Suggested revisions
\end{minipage}};

\draw[<->,line width=2pt](body) -- (a);
\draw[->,line width=2pt](body) -- (b);
\draw[->,line width=2pt](a) -- (c);
\draw[->,line width=2pt](b) -- (c);
\draw[->,line width=2pt](c) .. controls +(down:3cm) and +(right:5cm) .. (d);
\draw[->,line width=2pt](d) .. controls +(left:6cm) and +(down:3cm) .. (e);
\end{tikzpicture}
}
\end{center}
\caption{Iterative design driven by use cases.}
\label{myapproach}
\end{figure}

For a given translation of a normative framework $N$, the designer must specify what part of the theory is  subject to revision.
The  theory is split into two parts: a ``revisable'' part, $N_{T}$, and a ``fixed'' part, $N_{B}$. By default the former includes rules of the form (\ref{initiated}), (\ref{terminated}) and (\ref{occurred}), given in Figure~\ref{fig2}, and the latter includes the rest of the representation of the normative system and the set $T_{U}$ of the traces in $U$. 

 \begin{sloppypar}
{\em Given a set $U$ of use cases, a TR task for a normative framework $\mathcal{N}$ is defined as the tuple $\langle  O_{U}, N_{B} \cup T_{U}, N_{T}, M\rangle$,} where $M$ includes by default a body declaration for any static relation declared in $N_{B}$, and the following mode declarations (where the schema is opportunely formed by substituting arguments with input placemarkers):
$modeh(occurred(e^{*}, +instant))$, for each $e \in \instevents$;
$modeh(initiated(f^{*}, +instant))$ and $modeh(terminated(f^{*}, +instant))$, for each $f \in \fluents$;
$modeb(holdsat(f^{*}, +instant))$, for each $f \in \fluents$;
$modeb(occurred(e^{*}, +instant))$, for each $e \in \events$.
 \end{sloppypar}
%
%
%
%
The choice of the set of mode declaration $M$ is crucial and is ultimately the responsibility of the designer. Many mode declarations ensure higher coverage of the specification but increase the computation time. Conversely, fewer mode declarations improve performance but may result in partial solutions. The choice may be driven, for example, by previous design cycles, or interest in more problematic parts of the specification.

As shown in Figure \ref{myapproach} the design of a normative system is an iterative process. The representation $N$ in $\AnsProlog$ of a system described by the designer using a normative language is tested against a set of use cases also provided by the designer. This analysis step is performed by running an ASP solver over $N$, extended with the observed events included in the use cases, and a constraint indicating that no answer set that does not satisfy $O$ is acceptable. 
Conceptually, if the solver is not able to find an answer set (i.e. returns unsatisfiable), then some of the given use cases are not satisfied in the answer sets of $N$ and a revision step is performed. Possible revisions are provided to the designer who ultimately chooses the most appropriate one.

\subsection{Case Study}\label{sec:example}
We illustrate the methodology with a small but rich enough case study that demonstrates the key properties and benefits of our proposed approach. The following is a description of a reciprocal file sharing normative framework. 
\begin{quote}\small The active parties---agents---of the scenario find themselves
initially in the situation of having ownership of several (digital)
objects---the blocks---that form part of some larger composite
(digital) entity---a file.  An agent is required to share a copy of a block they hold before they can download a copy of block they are missing.  Initially each agent holds the only copy
of a given block and there is only one copy of each block in
the agent population.  
Some {\em vip} agents are able to download blocks without any restriction. Agents that request a download and have not shared a block after a previous download generate a violation for the download action and a misuse violation for the agent. A misuse terminates the empowerment of the agent to download blocks.\end{quote}


\noindent
The designer devises the following use case  $\langle T,
O\rangle$:

\begin{tabular}{ l r }
\begin{minipage}{2in}
{\scriptsize\begin{equation*}\hspace{-0.5cm}
T =   \left\{ \begin{array}{l} 
observed(start,i00).\\
observed(download(alice,bob,x3),i01).\\
observed(download(charlie,bob,x3),i02).\\
observed(download(bob,alice,x1),i03).\\
observed(download(charlie,alice,x1),i04).\\
observed(download(alice,charlie,x5),i05).\\
observed(download(alice,bob,x4),i06).
\end{array} \right.  \label{clause:firstrule:T}
\end{equation*}
}
\end{minipage}
&
\begin{minipage}{2.9in}
{\scriptsize
\begin{equation*}
O =   \left\{ \begin{array}{l} 
not\ viol(myDownload(alice,x3),i01).\\
not\ viol(myDownload(charlie,x3),i02).\\
not\ viol(myDownload(bob,x1),i03).\\
not\ viol(myDownload(charlie,x1),i04).\\
not\ viol(myDownload(alice,x5),i05).\\
viol(myDownload(alice,x4),i06).\\
\end{array} \right.  \label{clause:firstrule:O}
\end{equation*}

}
\end{minipage}
\end{tabular}

%
%
%
%
%
%
%
%
%
%
%
%
%
%
%
%
%
%
%
%
%

The use case models a sequence of events that includes a violation at the time point $i06$, while the $download$ events at the other time points do not generate violations. 
In the trace, $charlie$ performs a download at time point $i04$ without sharing a block after the last download. This is not expected to generate a violation since $charlie$ is defined as $vip$ ($isVIP(charlie) \in N$). 


The initial normative system includes the domain component and type definitions given in Figure \ref{second} and a specific component given by the following revisable theory $N_{T}$:

%

\begin{lstlisting}
%rule 1
initiated(hasblock(X,B),I) :- 			
	occurred(myDownload(X,B),I).
%rule 2
initiated(perm(myDownload(X,B)),I) :- 		
	occurred(myShare(X),I).
%rule 3
terminated(pow(extendedfilesharing,myDownload(X,B)),I) :- 
	occurred(misuse(X),I).
%rule 4
terminated(perm(myDownload(X,B2)),I) :- 
	occurred(myDownload(X,B),I).
%rule 5	
occurred(myDownload(X,B),I) :- 
	occurred(download(Y,Y,B),I), holdsat(hasblock(Y,B),I).
%rule 6	
occurred(myShare(X),I) :- 		
	occurred(download(Y,X,B),I), holdsat(hasblock(X,B),I).	
\end{lstlisting}

Given the use case and the above formalisation of the normative system, the first iteration of our approach proposes, through the revision process, the deletion of a condition in  {\em rule 5} and addition of a condition to {\em rule 4} as shown below (leaving the other rules unaltered):

\begin{lstlisting}
%rule 4 - revised
terminated(perm(myDownload(X,B2)),I) :- 
	not isVIP(X), occurred(myDownload(X,B),I).
%rule 5 - revised
occurred(myDownload(X,B),I) :- 
	holdsat(hasblock(Y,B),I).
\end{lstlisting}

However, this is not yet the intended formalisation. 
As an additional debugging facility the designer can request the set of violations that are true in the answer sets that corresponds to the revision and notice that unwanted violations are generated at each time point. This feedback can be used to refine the use case provided. In fact the use case specifies the single specific violations that must {\em not\/} occur but it does not request explicitly that no violations should occur in the first five time points (e.g.  {\small \em{viol(myDownload(alice,x3),i02), viol(myDownload(alice,x4),i02)}}). These violations can be observed in the answer set associated with the revision. The designer can then improve the use case by modifying the set of expected outputs:

{\scriptsize
\begin{equation*}
O =   \left\{ \begin{array}{l} 
viol(myDownload(alice,x4),i06).\\
not\ viol(myDownload(A,B),T),  T!= i06. \\
occurred(misuse(alice), i06). \\
not\ occurred(misuse(X),  T),  T!= i06. \\
\end{array} \right.  \label{clause:firstrule}
\end{equation*}
}


%
%
%
%
%
%
%
%

In the subsequent iteration, the revision process suggests changes that include those identified in the previous iteration (i.e. addition of condition in {\em rule 4} and deletion of condition in rule 5), and the addition of a further condition in the body of rule 5. The combined effect of these changes fixes the original error in the specification, by also changing the name of one of the variables. 
Furthermore, since the output $O$ of the use case includes a desired $misuse$ event, which is not currently formalised in the system, the revision also suggests the new rule 7 given below. The final theory $N_{T}'$ includes the following rules (leaving untouched rules 1, 2, 3 and 6)\footnote{The revision is generated in 23 seconds by \textsc{iclingo}\cite{Gebser07clasp:a} on a 2.8 GHz Intel Core 2 Duo iMac with 4 GB of RAM.}:

\begin{lstlisting}
%rule 4 - revised
terminated(perm(myDownload(X,B2)),I) :- 
	not isVIP(X), occurred(myDownload(X,B),I).
%rule 5 - revised	
occurred(myDownload(X,B),I) :- 
	occurred(download(X,Y,B),I),  holdsat(hasblock(Y,B),I).
%rule 7 - new
occurred(misuse(X),I) :- 
	occurred(viol(myDownload(X,B)),I).
\end{lstlisting}


In summary, after a few iterations {\em rule 4} is corrected by adding an exception $not$ {\em isVIP(X)},  {\em rule 5} is revised by correcting a typographical error in its condition (i.e. the name of a variable was not the intended one -- {\em occurred(download({\bf Y},Y,B),I)}), and finally, a new rule is learnt that defines $misuse$ coherently with respect to the provided use case.

\section{Theory revision through ASP}\label{sec:details}
In this section we provide more details about the revision process. We first introduce all the computational steps to derive a revision with respect to a set of use cases. Then we delve into the details of the learning system, describing the integrated ASP-based ILP approach.

The revised normative system $N_{B} \cup N_{T}'$ is computed by means of two program transformations and an abductive reasoning process executed in ASP, which derives prescriptions for revisions and new rules in the form of abducibles. The abductive solution has a one-to-one mapping to a revision of the initial theory.

\subsection{Revision}
The approach described in this section can be applied to other problems of TR. To the best of our knowledge, our methodology is the only one currently available that is able to support revision of non-monotonic \AnsProlog theories that supports integrity constraints, aggregates and other ASP constructs, providing revisions as answer sets.
Operationally, the revision is performed using a similar transformation to the one described in \cite{Corapi2009}. Figure~\ref{steps} details the revision steps for one of the rules in the case study described above and Algorithm~\ref{alg:algstep} illustrates the phases. We present the conceptual steps and refer the reader to \cite{Corapi2009} for further details.

\begin{algorithm}
\caption{Phases of the revision algorithm.}\small
\KwIn{$N_{B}$ fixed theory; $N_{T} \in s(M)$ revisable theory; $P$ set properties; $M$ mode declarations }
\KwOut{$N_{T}'$ revised theory according to the given $P$}
\hspace{2em}($\overline{N_{T}}, \overline{M}) =  \mbox{pre-processing}(N_{T}, M)$\;
\hspace{2em}$H = \mbox{ASPAL}(P, N_{B} \cup \overline{N_{T}},\overline{M})$\;
\hspace{2em}$N_{T}' = \mbox{post-processing}(N_{T}, H)$\;
\hspace{2em}return  $N_{T}'$\;
\label{alg:algstep} 
\end{algorithm}

A {\em pre-processing phase} lifts the standard ILP process of learning
hypotheses about examples up to the (meta-)process of 
learning hypothesis about the rules and their exception
cases. For every rule in $N_{T}$, every body literal $c^{i}_{j}$ is replaced by the atom
$try(i,j,c^{i}_{j})$, where $i$ is the index of the
rule, $j$ is the index of the body literal in the rule and
the third argument is a reified term for the literal
$c^{i}_{j}$. 
$not\ exception(i, h_{i}, v_{i})$ is added to the body of the rule
where $i$ is the index of the rule, $h_{i}$ is the
reified term for the head of the rule and $v_{i}$ is an optional list of additional variables appearing in the body (see Figure~\ref{steps}).
The $try$ predicate is defined in such a way that whenever $del(i, j)$ is true, the meta-condition $try(i,j,c^{i}_{j})$ is always true. Otherwise $try(i,j,c^{i}_{j})$ is true whenever $c^{i}_{j}$ is true. Facts of the type $del(i, j)$ can be learnt by the ILP system used within the revision.
$M$ specifies mode declaration of rules that can be added together with additional head declarations that are added to take into account the newly introduced $del$ and $exception$ predicates. 

\begin{figure}[!t]
\begin{multicols}{2}
{\em 1 -- Pre-processing (rules in $\overline{N_{T}}$)}
\begin{lstlisting}
terminated(perm(myDownload(X,B2)),I) :- 
  try(4, 1, occurred(myDownload(X,B),I)),
  not exception(terminated(perm(myDownload(X,B2)),I), B).

try(4, 1, occurred(myDownload(X,B),I)) :-
  not del(4, 1),
  occurred(myDownload(X,B),I).
  
try(4, 1, occurred(myDownload(X,B),I)) :-
  del(4, 1).
\end{lstlisting}
{\em 2 -- Learning (rule in H)}
\begin{lstlisting}
exception(terminated(perm(myDownload(X,B2)),I), B) :- 
  isVIP(X).
\end{lstlisting}
{\em 3 -- Postprocessing (rule in $N_{T}'$)}
\begin{lstlisting}
terminated(perm(myDownload(X,B2)),I) :- 
  not isVIP(X), 
  occurred(myDownload(X,B),I).
\end{lstlisting}
\end{multicols}
\caption{Detailed revision transformations for {\em rule 4}  (Section~\ref{sec:example})}
\vspace{-5mm}
\label{steps}
\end{figure}

In the {\em learning phase}, given the pre-processed theory $\overline{N_{T}}$ and the new mode declarations $\overline  M$, the following ILP task is executed $\langle P, N_{B} \cup \overline{N_{T}}, \overline M\rangle$, using \textsc{aspal}, the learning system described in Section \ref{sec:learning:aspal}.
The outcome of the learning phase $H$ is used in a {\em post-processing phase} which generates a revised
theory $N_{T}'$ semantically equivalent to  $\overline{N_{T}}\cup H$.
Informally, for each $del(i, j)$
fact in $H$ the corresponding condition $j$ in rule $i$ in $N_{T}$
is deleted. 
For each exception rule in $H$ of the form
$exception(i,
h_i,v_{i})\ifl c_{1}, ..., c_{n}$, the corresponding rule
$i$ in $N_{T}$ is substituted with $n$ new rules, one for each
condition $c_{h}$, $1\leq
k\leq n$. Each of these rules $k$ will have in the head
the predicate $h_{i}$ and in the body all conditions present
in the original rule $i$ in $N_{T}$ plus the
additional condition $not\ c(k)$. An exception with empty
body results in the original rule $i$ being deleted. An exception for which at least two conditions
share variables is kept as an additional ``exception concept'' in the revised theory. 
The pre-processing and post-processing phases perform syntactic transformations that are answer set preserving and do not involve the answer set solver.



\subsection{ASPAL }\label{sec:learning:aspal}


The system used in this work, called \textsc{aspal} (ASP Abductive Learning),  
though used here to support the revision of a normative system,
can be applied more generally to non-monotonic ILP problems.
It is based on the transformation from an ILP task to an abductive reasoning task, used in a recently proposed ILP system \cite{corapi_et_al}.

This system offers several advantages over other existing ILP approaches, making it particularly suited for normative design. \textsc{aspal} is able to handle negation within the learning process, and therefore reason about default assumptions governing inertial fluents; to perform non-observational and multiple predicate learning, thus computing hypotheses about causal dependencies between observed sequences of events and normative states; and to learn non-monotonic hypotheses, which is also essential for theory revision. Furthermore, the learning can be enabled by a simple transformation of the mode declarations and does not require the computation of a {\em bridge theory} \cite{yamamoto}.
As discussed in \cite{corapi_et_al}, none of the existing ILP systems provides the above mentioned features. 
Embedding the learning process within ASP reduces the semantic gap between the normative system and the learning process and permits an easier control of the whole process. The notion of revision distance as in Definition \ref{tr1} can be managed by the optimisation facilities provided by modern ASP solvers \cite{Gebser07clasp:a}. Optimisation statements can be used to derive answer sets that contain a minimal number of atoms of a certain type that ultimately relate to new rules or revisions, as explained in this section.


%
%
%
%
%
%

As in \cite{corapi_et_al}, an ILP task  $\langle P, B, M\rangle$ is transformed into an abductive logic programming problem \cite{DBLP:journals/logcom/KakasKT92}, thus enabling the use of $\AnsProlog$.
Let us introduce some preliminary notation. Given a mode declaration $modeh(s)$ or $modeb(s)$, $id$ is a unique identifier for the mode declaration, ${\bf s}$ is the literal obtained from $s$ by replacing all placemarkers with different variables $X_{1}, ...,X_{n}$; $type({\bf s}, s)$ denotes the sequence of literals $t_{1}(X_{1}), ...,t_{n}(X_{n})$ such that $t_{i}$ is the type of the placemarker replaced by the variable $X_{i}$; 
$con({\bf s}, s) = (C_{1}, ...,C_{c})$ is the {\em constant list} of variables in  $\bf{s}$ that replace only constant placemarkers in $s$. $inp({\bf s}, s) = (I_{1}, ...,I_{i})$ and $out({\bf s}, s) = (O_{1}, ...,O_{o})$ are defined similarly for input and output placemarkers.  Since $s$ is clear from the context, in the following we omit the second argument from $type({\bf s}, s)$, $con({\bf s}, s)$, $inp({\bf s}, s)$ and $out({\bf s}, s)$.

Given a set of mode declarations $M$, a {\em top theory} $\top = t(M)$ is constructed as follows:
\begin{itemize}\small
\item For each head declaration $modeh(s)$, with unique identifier $id$, the following rule is in $\top$

{\begin{equation} \label{head1} 
\begin{array}{ll} 
{\bf s} \ifl \\
\hspace{0.8em} rule(RId, (id, con({\bf s}), ()), \\
\hspace{0.8em} rule\_id(RId), \\
\hspace{0.8em} type({\bf s}), \\
\hspace{0.8em} body(RId, 1, inp({\bf s}))\\
\end{array} 
\end{equation}
}

\item For each body declaration $modeb(s)$, with unique identifier $id$ the following clause is in $\top$

{\begin{equation} \label{body1}
\begin{array}{ll} 
body(RId, L, I)   \ifl \\
\hspace{0.8em} rule(RId, L, (id, con({\bf s}), Links)), \\
\hspace{0.8em} link(inp({\bf s}), I, Link), \\
\hspace{0.8em}  {\bf s}, \\
\hspace{0.8em} type({\bf s}), \\
\hspace{0.8em}  append(I, out({\bf s}), O), \\
\hspace{0.8em} body(RId, L + 1, O) \\
\end{array}
\end{equation}}

\item The following rule is in $\top$ together with the definitions for the $link$, $rule\_id$ and $append$ predicates:

{\begin{equation}  \label{body2}
\begin{array}{ll} 
body(RId, L, \_)   \ifl rule(RId, L, last) \\
\end{array}  \notag
\end{equation}}

\end{itemize}

\begin{sloppypar}
\noindent$rule\_id(rid)$ is true whenever $1\! \leq\!rid\! \leq\!rn$ where $rn$ is the maximum number of new rules allowed. 
$link((a_{1}, ..., a_{m}), (b_{1}, ..., b_{n}), (o_{1}, ..., o_{m}))$ is true if for each element in the first list $a_{i}$, there exists an element in the second list $b_{j}$ such that $a_{i}$ unifies with $b_{j}$ and $o_{i} = j$. 
Given the top theory, we seek a set of $rule$ atoms $\Delta$, such that $P$ is true all models of $B \cup \top \cup \Delta$.
$\Delta$ has a one-to-one mapping to a set of rules $H = u(\Delta, M)$. Intuitively, each abduced atom represents a literal of the rule labelled by the first argument. The second argument collects the constant used in the literal and the third disambiguates the variable linking. Fig.~\ref{steps:2} shows the learning steps for {\em rule 4} of our example.  

For space limitations we only state the main soundness and completeness theorem \cite{aspalproof} of the learning system.\\

\begin{theorem}\label{theorem}
Given an ILP task $\langle P, B, M\rangle$, $H$ is an inductive solution if and only if there is a $\Delta$ such that $H = u(\Delta, M)$, $\top = t(M)$ and $P$ is true in all the answer sets of $B \cup \top \cup \Delta$.
\end{theorem}

The ASP solver is used to compute a set of solutions $\Delta$, that can be translated back into a set of inductive solution.
Soundness and completeness for the revision procedure rely on Theorem \ref{theorem} and on the underlying ASP solver properties. 
These properties also ensure that if a set of theories that matches the requirements exists within the language bias of the learning, in the limit, if a complete set of all use cases (an extensional specification of the requirements) is provided, the revision converges to the expected theory. 
This is of course an ideal case. In practice the system outputs more accurate solutions as more comprehensive use case sets are provided.

\end{sloppypar}

\begin{figure}[!t]
\begin{multicols}{2}
{\bf Inputs}\\
 \vspace{4mm}
{\em Mode declarations $M$}
\begin{lstlisting}
exception(terminated(perm(myDownload
	(+agent,+block)),+instant), +block).
\end{lstlisting}
 \vspace{4mm}
{\em Properties $P$}
 \vspace{2mm}
\begin{lstlisting}
viol(myDownload(alice,x4),i06).
not viol(myDownload(A,B),T),  T!= i06. 
occurred(misuse(alice), i06). 
not occurred(misuse(X),  T),  T!= i06. 
\end{lstlisting}
 \vspace{4mm}
{\em Background theory $B$}
 \vspace{2mm}
\begin{lstlisting}
terminated(perm(myDownload(X,B2)),I) :- 
  try(4, 1, occurred(myDownload(X,B),I)),
  not exception(terminated(perm(myDownload(X,B2)),I), B).

try(4, 1, occurred(myDownload(X,B),I)) :-
  not del(4, 1),
  occurred(myDownload(X,B),I).
  
try(4, 1, occurred(myDownload(X,B),I)) :-
  del(4, 1).
  
\end{lstlisting}
 \vspace{6mm}
\medskip
Top theory $\top$
\begin{lstlisting}
exception(4, terminated(perm(myDownload(A,B)),T)) :-
  instant(T),  block(B),  agent(A), 
  rule_id(RID),
  rule(RID, 0, (e4, (), ())),
  body(RID, 1, (A, B, T)).	

body(RID, Level, (A, B, T)) :-
  agent(A), block(B), instant(T),	
  rule_id(RID),
  link(L1, (A, B, T), LR1),
  rule(RID, Level, (isv, (), (LR1))),
  isVIP(L1),
  body(L + 1, RID, (A, B, T)).	

body(RID, L, _):-
  rule(RID, L, last).		
\end{lstlisting}
\hrule
\medskip
Abductive solution $\Delta$
\begin{lstlisting}
rule(0, 0, (e4, (), ())), 
rule(0, 1, (isv, (), (1))), 
rule(0, 2, last)
\end{lstlisting}
\hrule
\medskip
{\bf Output}\\
Inductive solution $H$
\begin{lstlisting}
exception(terminated(perm(myDownload
	(X,B2)),I), B) :- 
  isVIP(X).
\end{lstlisting}
\end{multicols}

\caption{Learning steps for {\em rule 4}  (Sec. \ref{sec:example}). We show only the relevant mode declarations and rules.}
\label{steps:2}
\end{figure}


\section{Discussion and Related Work}\label{sec:related}

The motivation behind this paper is the problem of how to converge
upon a complete and correct normative system {\em with respect
  to the intended range of application}, where in practice these
properties may be manifested by incorrect or unexpected behaviour in
use.  Additionally, we observe, from practical experience with
our particular framework, that it is often desirable 
to be able to develop and test
incrementally and regressively rather than attempt verification
once the system is (notionally) complete.

\begin{sloppypar}
The literature seems to fall broadly into three categories:
\begin{inparaenum}[(a)]
\item concrete language frameworks (OMASE
  \cite{DBLP:conf/aose/Garcia-OjedaDROV07}, Operetta
  \cite{DBLP:conf/atal/OkouyaD08}, InstAL \cite{cdvp06a}, MOISE
  \cite{DBLP:journals/ijaose/HubnerSB07}, Islander
  \cite{DBLP:conf/atal/EstevaCS02}, OCeAN
  \cite{DBLP:journals/ail/FornaraVVC08} and the constraint approach of
  Garcia-Camino et al. \cite{garcia-camino-et-al:2009}) for
  the specification of normative systems, that are typically supported
  by some form of model-checking, and in some cases allow for change
  in the normative structure;
\item logical formalisms, such as \cite{garion_et_al:DSP:2009:1904},
  that capture consistency and completeness via modalities and other
  formalisms like \cite{DBLP:conf/atal/BoellaPT09}, that capture the
  concept of norm change, or \cite{vasconcelos-et-al:2007} and
  \cite{cardoso+oliveira:2008};
\item mechanisms that look out for (new) conventions and handle their
  assimilation into the normative framework over time and subject to
  the current normative state and the position of other agents
  \cite{DBLP:conf/atal/Artikis09,DBLP:conf/atal/ChristelisR09}.
\end{inparaenum}
Essentially, the objective of each of the above is to realize a
transformation of the normative framework to accommodate some form of
shortcoming.  These shortcomings can be identified in several ways:
\begin{inparaenum}[(a)]
\item by observing that a particular state is rarely achieved, which
  can indicate there is insufficient normative guidance for
  participants, or
\item a norm conflict occurs, such that an agent is unable to act
  consistently under the governing norms \cite{kollingbaum-et-al:2007}, or
\item a particular violation occurs frequently, which may indicate
  that the violation conflicts with an effective course of action that
  agents prefer to take, the penalty notwithstanding.
\end{inparaenum}
All of these can be viewed as characterising emergent
\cite{savarimuthu_et_al:DSP:2009:1905} approaches to the evolution of
normative frameworks, where some mechanism, either in the framework,
or in the environment, is used to revise the norms.  In the approach
taken here, the designer presents use cases that effectively capture
the behavioural requirements for the system, in order to `fix' bad
states.  This has an interesting parallel with the scheme put forward
by Serrano and Saugar \cite{serrano+saugar:2010}, where they propose
the specification of incomplete theories and their management through
incomplete normative states identified as ``pending''.  
\end{sloppypar}

In \cite{boella_et_al:DSP:2009:1902}, whether the norms here are
`strong' or `weak' ---the first guideline--- depends on whether the
purpose of the normative model is to develop the system specification
or additionally to provide an explicit representation for run-time
reference.  Likewise, in respect of the remaining guidelines, it all
depends on how the framework is actually used: we
have chosen, for the purpose of this presentation, to stage norm
refinement so that it is an off-line (in the sense of prior to
deployment) process, while much of the discussion in
\cite{boella_et_al:DSP:2009:1902} addresses run-time issues.  Whether
the process we have outlined here could effectively be a means for
on-line mechanism design, is something we have yet to explore.
Within the context of software engineering, \cite{Alrajeh2006} shows how examples of desirable 
and undesirable behaviour of a software system can be used by an ILP
system, together with an incomplete background knowledge of the
envisioned system and its environment, to compute missing requirements
specifications. There are several elements in common with the scheme proposed here.


From an ILP perspective, we employ a system that can learn logic
programs with negation (stratified or otherwise) and, unlike other existing nonmonotonic ILP systems \cite{Sakama_nonmonotonicinductive}
is supported by completeness results, is integrated into ASP and can be tailored to particular design requirements.
Some properties and results of ILP in the context of ASP are shown in \cite{Sakama_learningby}. The author also proposes an algorithm for learning that is sound but not complete and, differently from the approach proposed here, employs a covering loop approach. 


\section{Conclusions and Future Work}


%
%

The motivation for this work stems from a real need for tool support
in the design of normative frameworks, because, although high-level, 
it is nevertheless hard for humans to identify errors in
specifications, or indeed to propose the most appropriate corrective
actions.  We have described a methodology for the revision of
normative frameworks and how to use tools with formal underpinnings to
support the process.  Specifically, we are able to revise a formal
model---represented as a logic program---that captures the rules of a
normative system.  The revision is achieved by means of inductive
logic programming, working with the same representation, informed by
use cases that describe instances of expected behaviour of the
normative system.  If actual behaviour does not coincide with
expected, theory revision proposes new rules, or modifications of
existing rules, for the normative framework.  Furthermore, given
correct traces, the learning process guarantees convergence---the
property of ``learning in the limit''.

From this firm foundation, which properly connects a theory of
normative systems with a practical representation, there are three
directions that we aim to pursue:
\begin{inparaenum}[(i)]
\item definition of criteria for selecting solutions from alternative
  suggestions provided by the learning (we are currently investigating the use of {\em crucial literals} \cite{Sattar:1991:UCL:116024.116026})
\item introduction of levels of confidence in the use cases and their
  use for selecting the ``most likely'' revision, in addition
  to the general criteria of minimal revision: i.e. combine some
  domain-independent heuristics with some domain-specific heuristics
  such as level of confidence in use cases
\item extension to interactions between normative frameworks and a
  form of cooperative revision.
\end{inparaenum}
Additionally, there is the matter of scalability.  The computation
time increases with the number of rules, time steps, errors in the
theory and in particular, mode declarations and language bias for the
learning.  That is, it grows with the state space of the
normative framework and the ``learning space'', i.e. is all possible
theories we can construct given our language bias.  We need to
experiment further to understand better to which factors performance
is sensitive and how to address these issues.

%
%
%
%

\bibliography{learning}
\label{lastpage}

\end{document}